\pdfoutput=1
\documentclass{crasty}
\usepackage{cite}
\usepackage{float}
\def\gsim{\mathrel{\rlap{\lower4pt\hbox{\hskip1pt$\sim$}}}}

\usepackage[]{lineno}
\begin{document}
%
%  Titel Page
%
\noindent
%DRAFT 1.0\\
LHeC-Note-2012-004 GEN \\
Geneva, August 13, 2012
%\today \\
%
\begin{figure}[h]
\vspace{-2.cm}
\hspace{13.5cm}
\includegraphics[clip=,width=.15\textwidth]{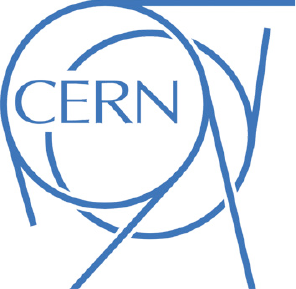}
\end{figure}
\begin{figure}[h]
\vspace{-1.3cm}
\hspace{4.3cm}
\includegraphics[clip=,width=0.45\textwidth]{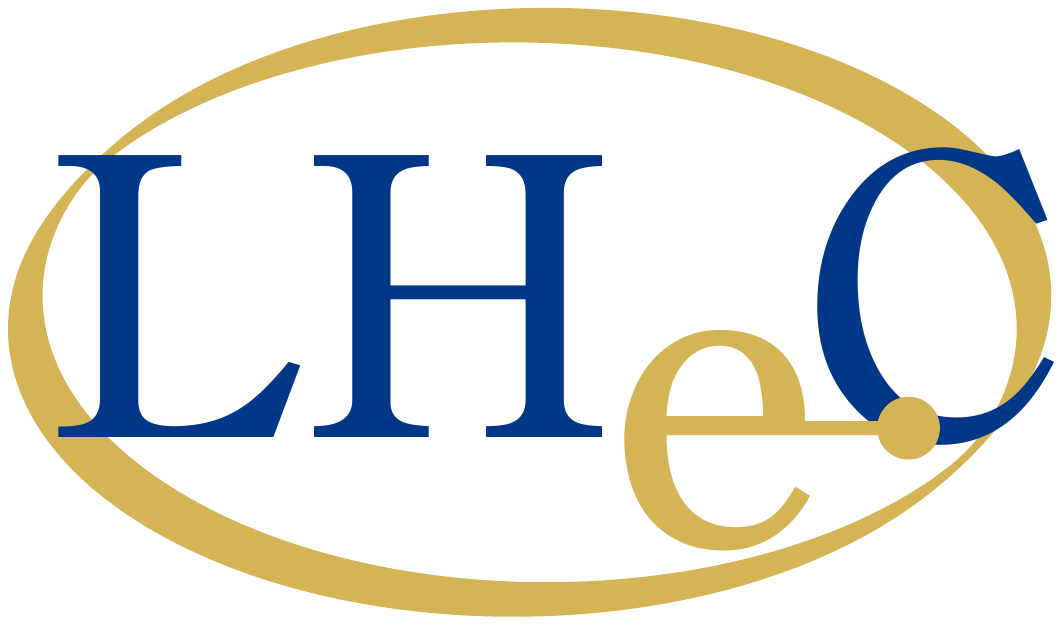}
\end{figure}
\begin{center}
\vspace{-.5cm}
%\vspace{4cm}
\begin{LARGE}
\bf{A Large Hadron Electron Collider  
at CERN} \\
\end{LARGE}
\vspace{0.5cm}
\begin{LARGE}
% Report on the Physics and Design \\
%Concepts for Machine and Detector \\
 \end{LARGE}
 \vspace{1cm}
\begin{Large}
\bf{LHeC Study Group
\footnote{For the current author list
see the end of this note. Contacts: 
oliver.bruning@cern.ch and max.klein@cern.ch}
%\footnote{
%The study group includes basically the authors of the
%deign report recently published  and a number of
%colleagues, who recently joined the project. 
%have recently expressed an interest to join the LHeC
%development. 
%The authorlist is available in the web version of this note.}} 
}\\ 
\end{Large}
\vspace{0.5cm}
$\bf{Abstract}$ \\
\vspace{0.3cm}
\end{center}
This document provides a brief overview of
the recently published report on the design
of the Large Hadron Electron Collider (LHeC), which comprises
its physics programme, accelerator physics, technology
and main detector concepts.
The LHeC exploits and develops
challenging, though principally existing, accelerator and detector
technologies.
This summary is complemented by brief illustrations of some of
the highlights of the physics programme,
which relies on  a vastly extended kinematic
range, luminosity and unprecedented precision in
deep inelastic scattering. Illustrations
are provided regarding high precision QCD, 
new physics (Higgs, SUSY)
and electron-ion physics. The LHeC is designed to
run synchronously with the LHC in the twenties 
and to achieve an integrated luminosity of O(100)\,fb$^{-1}$.
It will become the cleanest high resolution microscope
of mankind and will substantially extend as well as complement
the investigation of the  physics of the TeV energy scale,
which has been enabled by the LHC. 
\\
\vspace{1cm}
\begin{center}
Submitted to the Cracow Symposium of the \\
Update of the European Strategy for Particle Physics, \\
Cracow (Poland) September 2012.
%THIS IS A DRAFT VERSION OF THE LHeC CONTRIBUTION WHICH WILL BE REPLACED BY THE FINAL %VERSION ON AUGUST 8$^{th}$, 2012
\end{center}
%
%\newpage
%\noindent \input{cracauthors}
\newpage
%
% ---------------------------------------------------------------------------------
%
\section{Introduction - the LHeC on one page}
Deep inelastic lepton-hadron scattering (DIS)  represents the cleanest probe
of partonic behaviour in protons and nuclei. Highest energy electron-parton
collisions provide unique information on the physics beyond the Standard Model (SM).
% allow new particles with both lepton and baryon quantum numbers,
%as predicted in various theories, to
%be singly produced with a high cross section. 
The principal aim of the LHeC conceptual design 
report (CDR)~\cite{AbelleiraFernandez:2012cc}
has been to lay out the design concepts for a second generation DIS
electron-proton ($ep$), and a first electron-ion ($eA$) collider, 
taking unique advantage of 
the intense, high energy hadron beams of the Large Hadron Collider.
The LHeC in its default design configuration uses a $60$\,GeV
electron beam of high intensity based on a racetrack, energy recovery 
configuration using two  $10$\,GeV electron linacs.
It thus exceeds the luminosity of HERA by a factor of $100$
and reaches a maximum $Q^2$ of above $1$\,TeV$^2$ as compared
with a maximum of $0.03$\,TeV$^2$ at HERA.  
Correspondingly the lowest Bjorken $x$  covered in
the DIS region with the LHeC is about $10^{-6}$, where gluon saturation is
expected to exist. This coverage allows a multitude of crucial
DIS measurements to be performed, to complement and extend the 
search potential for new physics at the LHC, and
it  also makes the LHeC a testing ground for the Higgs boson cleanly
produced in $WW$ and $ZZ$ fusion in $ep$. The extension of the kinematic
coverage in DIS lepton-ion collisions amounts to nearly $4$ orders of magnitude
and can be expected to completely change the understanding
of quark-gluon interactions in nuclei, tightly constraining the initial conditions
of the formation of the quark-gluon plasma (QGP).
The LHeC project represents a unique possibility to take forward 
the field of DIS physics as an integral part of the future high
energy physics programme. It enhances the exploration of the accelerator
energy frontier with the LHC.
Naturally it is linked to the LHC time schedule
and lifetime, which is estimated to continue for two decades hence. 
Therefore, a design concept has been presented
which uses available, yet challenging, technology, both for the
accelerator and for the detector, and time schedules are considered 
for realising the LHeC within about the next decade.

The CDR~\cite{AbelleiraFernandez:2012cc} describes in considerable detail
two options for the LHeC, a ring-ring (RR) and a linac-ring (LR) 
configuration. In a recent workshop~\cite{chavannes12} following the
publication of the CDR, it was decided to pursue the technical
design work for the LR configuration only, keeping the RR as a backup in
case new developments suggest to come back to it or if the LR
design meets insurmountable problems.  For comparison,
the main parameters for both the RR and the LR configurations
 are listed in  Table\,\ref{tabpar}.
The luminosity is constrained by a chosen wall-plug power limit of 
$100$\,MW for the lepton beam. The actual $e$ beam power 
consumption  is therefore limited to a few tens of MW. 
The linac option, however, effectively uses almost a GW
of beam power by recovering the energy of the spent beam.
Both the RR and the LR option are designed to provide 
$10^{33}$\,cm$^{-2}$s$^{-1}$ luminosities.
The LR configuration has high electron beam polarisations but
realistically has a significantly reduced $e^+p$ luminosity
with respect to $e^-p$. To mitigate this limitation,
various R\&D options are
presented in the CDR.  The LHeC parameters rely on the so-called ultimate LHC
beam configuration. From today's experience with the LHC operation,
even more performant proton beam parameters can be expected. It is thus
possible that the improved proton beam parameters 
of the HL-LHC upgrade will lead to a significantly 
higher luminosity for the LHeC than is quoted here.
The first estimates of the  luminosity in $eA$
point to a good basis for low $x$
electron-ion scattering measurements, even in time-restricted 
periods of operation. More refined studies are required, in particular
for the case of deuterons, which have yet to be used in the LHC.
The small beam spot size is  particularly well suited
for tagging of charm and beauty decays.
Backscattered laser techniques 
can provide a real photon beam derived from the $e$ linac
with rather high efficiency, which
would give access to $\gamma p$ and $\gamma A$ 
or even $\gamma \gamma$ physics at high energies.
\begin{table}[hbt]
   \centering
   \begin{tabular}{|l|c|c|}
%       \hline
%     &  Ring   & Linac \\
       \hline
      electron beam $60$ GeV & Ring  & Linac \\
\hline
%beam energy $E_e$ & \multicolumn{2}{c}{$60 \ \rm{GeV}$}   \\
%beam energy $E_e$ $\rm{GeV}$ & 60 & \\
$e^-$ ($e^+$)  per bunch $N_e$ [$10^{9}$]  &  $ 20~(20) $ & $ 1~(0.1)$   \\
$e^-$ ($e^+$) polarisation [\%]& $40~(40)$ & $90~(0)$ \\
bunch length [mm]  &  $ 6$ & $ 0.6$ \\
tr. emittance at IP $\gamma \epsilon^e_{x,y}$ [ mm] & $ 0.59,~0.29 $ & $ 0.05 $ \\
IP $\beta$ function $\beta^*_{x,y}$ [m] & $ 0.4,~0.2 $ & $ 0.12 $  \\
beam current [mA] & $ 100 $ & $ 6.6 $ \\
energy recovery efficiency [\%] & $ - $ & $ 94 $ \\
%total wall plug power [MW] & $100$ & $100$    \\
%syn rad power [$\rm{kW}$] &  $ 51 $ & $ 49 $ \\
%critical energy [$\rm{keV}$] &  $ 163 $ & $ 718$ \\
\hline
       proton beam 7 TeV & & \\
       \hline
%beam energy $E_p$ & \multicolumn{2}{c}{$7 \ \rm{TeV}$}        \\
protons per bunch $N_p$ [$10^{11}$] & $1.7$ & $1.7$        \\
transverse emittance $\gamma \epsilon^p_{x,y}$ [$\rm{\mu m}$] & $3.75$ & $3.75$   \\
\hline
      collider & & \\
\hline
%luminosity $10^{\circ}$ [$10^{32}$cm$^{-2}$s$^{-1}$] & $ 18 $ & $ - $\\
Lum $e^-p$ ($e^+p$) [$10^{32}$cm$^{-2}$s$^{-1}$] & $ 9~(9) $ & $ 10~(1) $\\
bunch spacing [$\rm{ns}$]& $25$ & $25$   \\
rms beam spot size $\sigma_{x,y}$ [$\rm{\mu m}$] & $ 45,22 $ & $ 7 $\\
crossing angle $\theta$ [mrad] & $ 1 $ & $ 0 $\\
$L_{eN}=A~L_{eA}$ [$10^{32}$cm$^{-2}$s$^{-1}$] & $ 0.45 $ & $ 1 $\\
\hline
   \end{tabular}
   \caption{
\footnotesize{
Baseline design parameters of the Ring (RR)
 and the Linac (LR) configurations of the LHeC.  The technical design
will be pursued for the electron Linac with the Ring as a 
back-up configuration. The luminosity corresponds to a total
wall plug power limit of $100$\,MW for the electron beam. The LHeC physics
programme primarily uses protons but foresees also 
electron collisions with heavy ions ($eA$) and
with deuterons ($eD$).}
}
   \label{tabpar}
\end{table}

Section 2 provides first an extended overview of the LHeC,
its physics potential, further details
on the accelerator and a suitable detector concept.
The section concludes with remarks on the time schedule and on
synergies with related developments. In Section 3 
the physics programme is illustrated, with  
a brief physics overview table followed by
one-page descriptions of selected key topics: PDFs, low $x$
and heavy ion physics, supplemented by discussions on the
possible importance of the LHeC for the study of the
Higgs boson and for searches for supersymmetry (SUSY).
The paper concludes with a short comment on the LHeC status
and next steps. It relies on the CDR~\cite{AbelleiraFernandez:2012cc}
and novel physics, technical
and organisational considerations as were presented at the
June 2012 CERN-ECFA-NuPECC workshop~\cite{chavannes12} 
on the LHeC. The current text replaces the 
draft version of the  paper submitted on July 31, 2012.

\section{An Overview of the LHeC}
\subsection{Physics}
\subsubsection*{Cornerstones of the Physics Programme}
The LHeC, with a multi-purpose detector, has a unique
physics programme of deep inelastic scattering,
which can be pursued with unprecedented precision over a hugely
extended kinematic range.
This comprises a per mille level precision measurement of $\alpha_s$,
accompanied by  ultra-precise charm and beauty density measurements,
 the accurate mapping of the gluon field over five orders of magnitude in
Bjorken $x$, from $x \simeq 3 \cdot 10^{-6}$ up to $x$ close to $1$, 
% the unbiased resolution of the complete quark content of the nucleon,
the first assumption-free, all-flavour PDFs
including first direct measurements of the $Q^2$ and $x$ dependences
of the strange and top quark distributions, and
the resolution of the partonic structure of the photon.
Neutron and nuclear structure can be explored 
in a vast new kinematic region, as these were uncovered by HERA, and 
high precision electroweak measurements can be made, for
example of the scale dependence of the weak mixing angle $\sin^2 \Theta_W$ and 
of the light-quark weak neutral current couplings. These and 
more exclusive measurements of e.g.
jets and diffraction at high energy and mass scales, represent new challenges
for the development of Quantum Chromodynamics to a new level of precision.
By accessing very low $x$ values, down to $10^{-6}$ 
at $Q^2 \simeq 1$\,GeV$^2$, the LHeC is expected to resolve the 
question of whether and how partons exhibit non-linear interaction dynamics
where their density is particularly high, and whether indeed there
is a damping of the rise of the parton densities towards low $x$,
a question also related to ultra-high energy neutrino physics, which
probes $x$ values as small as $10^{-8}$.
\subsubsection*{Relation to QCD: Developments and Discoveries}
The ultra-high precision measurements with the LHeC
challenge perturbative QCD to be further developed, by 
preparing for a consistent DIS analysis to N$^3$LO.
Precision measurements of generalised parton distributions in DVCS
are necessary for the development of a parton model theory
based on scattering amplitudes and the development of a
3-dimensional view of the proton. Analysis in the extended phase space
will pin down the mechanism of parton emission and will
determine unintegrated, transverse momentum dependent
parton distributions for the description of $ep$ as well as $pp$
final states. The coverage of  extremely
low $x$ regions at $Q^2 \geq 1$\,GeV$^2$,
both in $ep$ and in $eA$, will establish the basis for the
theoretical development of  non-linear parton evolution physics.
High energy $ep$ scattering 
may be important for constructing a non-perturbative approach to
QCD based on effective string theory in higher dimensions.
Instantons are a basic aspect of non-perturbative QCD,
which also predicts the existence of the Odderon,
a dressed three-gluon state,
and both are yet to be discovered.
A new chapter in $eA$ scattering will be opened
with measurements of unprecedented kinematic range and precision,
 allowing huge progress in the understanding of
partonic interactions in nuclei, which is still in its infancy.
It will also probe
the difference between hadronisation phenomena inside and
outside the nuclear medium. The establishment of an
ultra-high parton density,
``black-body'' limit in DIS would change the scaling behaviour
of the structure functions and the rates with which diffraction
and exclusive vector meson production occur. QCD is a subtle theory
which is far from being mastered and many of its areas call
for a renewed and extended experimental basis.
\subsubsection*{Relations to LHC Physics}
Deep inelastic scattering is the ideal process for the determination
of the quark and gluon distributions in the proton.
Studies of the parton
substructure of the nucleon are of great interest for the development 
of strong interaction theory, 
but they are also a necessary input for new
physics searches and studies at the LHC, whose potential will be
correspondingly enhanced. 
With the increasingly
apparent need to cover  higher and higher new particle
masses in this
endeavour, it becomes ever more important to
pin down the parton behaviour at large $x$,
which governs both signal and background rates near to the
LHC kinematic limit.
An example is the prediction of gluino pair production cross sections
from gluon-gluon fusion, which are currently not well known 
at masses beyond a few TeV, and for which a new level of precision
on the gluon distribution will be critical.
Similar situations are expected to arise
in future studies of
electroweak and other new physics, where
large $x$ parton distributions will play a crucial role.
QCD predicts factorisation as well as
resummation phenomena which can be
tested with much enhanced sensitivity
by combining LHC and LHeC results in inclusive and
also in diffractive scattering. Certain
parton distribution constraints, e.g. for the strange quark,
are also derived from Drell-Yan measurements of $W$ and
$Z$ production at the LHC, which will be verified with much
extended range, accuracy and completeness at the LHeC.
The $eA$ measurements determine the parton densities and 
interaction dynamics
in nuclei and are therefore a natural and necessary complement to
the $AA$ and $pA$ investigations made with the LHC.

Depending on what new phenomena are found at the LHC,
which has a superior cms energy compared to the LHeC
(and to any of the proposed $e^+e^-$ colliders), there are various
scenarios where the cleaner $ep$ initial state can help substantially
to clarify and to investigate new physics.  Key examples are
the spectroscopy of leptoquarks,
$R$-parity violating SUSY states, 
substructure and contact interaction phenomena,
as well as the search for excited electron or neutrino states. 

The Higgs particle is produced in $WW$ and $ZZ$ fusion
in $ep$ collisions at the LHeC. 
These production modes can be uniquely identified
by the nature of the charged or neutral current process, and
decays can be studied with low background, including the
dominant decay to $b \overline{b}$ to about $4$\,\% precision. 
From the $WW$ production the contributions
from CP even (SM) or odd (non-SM) 
Higgs quantum numbers can be unfolded. 

As the LHC results continue to appear and
the LHeC design proceeds, the relation between the two projects
will become a more central part of the developments
of the physics, the detector and the machine. 

\subsection{Accelerator}
%\subsection{Linac Design and Beam Dynamics}
%\input{daniel}
\subsubsection*{Electron Beam Layout and Civil Engineering}
\begin{figure}
%\centerline{\includegraphics[angle=0,clip=,width=1.\textwidth]{LHeCview}}
\centerline{\includegraphics[angle=0,clip=,width=.9\textwidth]{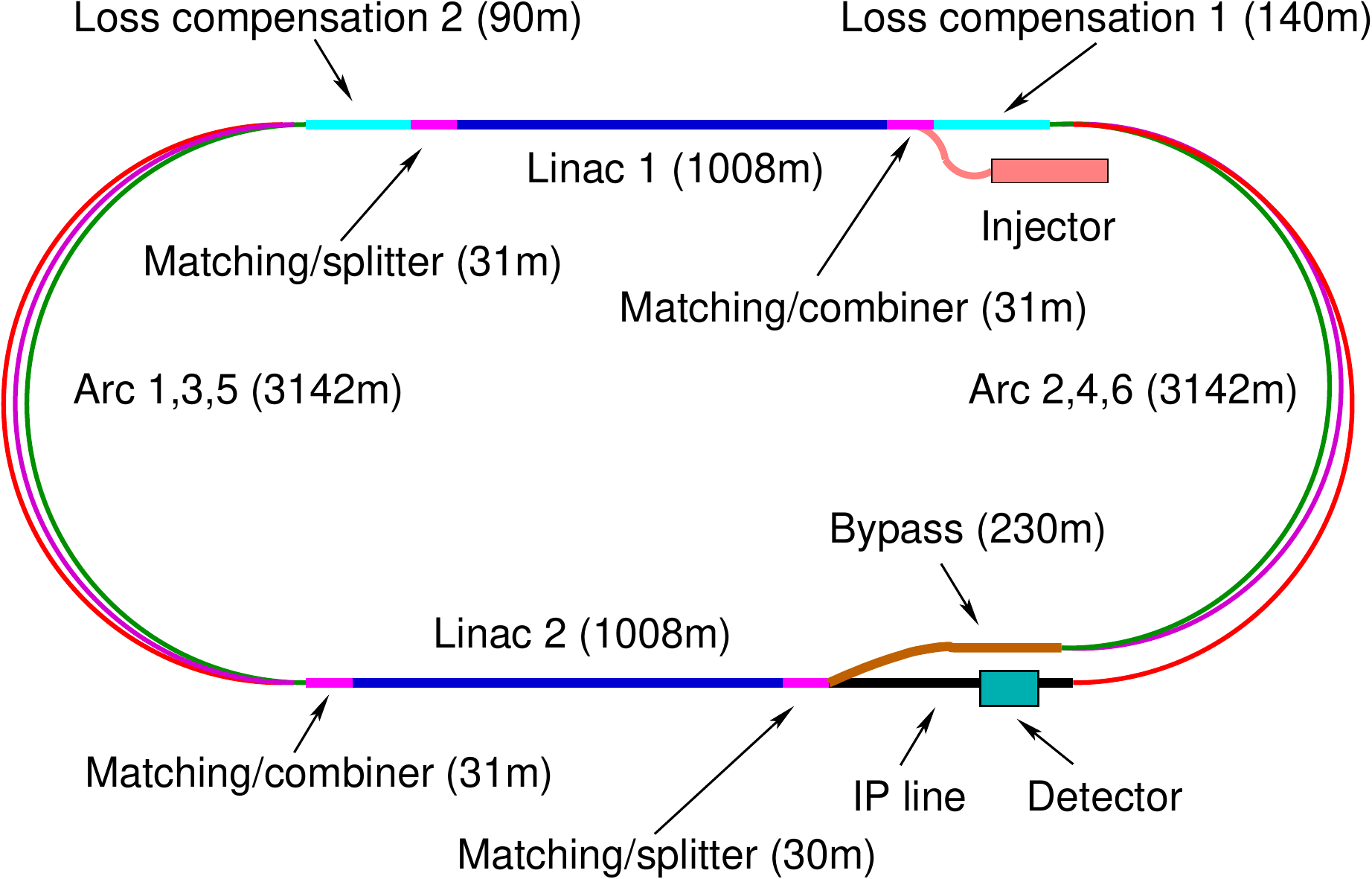}}
%\centerline{\includegraphics[angle=0,clip=,width=.9\textwidth]{ERLlayout}}
\caption{
\footnotesize{
Schematic view on the LHeC racetrack configuration.
Each linac accelerates the beam to $10$\,GeV, which leads to a
$60$\,GeV electron energy at the collision point with three
passes through the opposite linear structures of $60$ cavity-cryo modules
each. The arc radius is about $1$\,km, mainly determined
by the synchrotron radiation loss of the $60$\,GeV beam which 
is returned from the IP and decelerated for recovering the beam power.
Comprehensive design studies
of the lattice, optics, beam (beam) dynamics, dump, IR and
return arc magnets, as well as
auxiliary systems such as RF, cryogenics or spin rotators 
are contained in the CDR \cite{AbelleiraFernandez:2012cc},
which as for physics and detector had been reviewed by 
24 referees appointed by CERN.
}}
\label{fig:liview}
\end{figure}

The default electron beam energy is set to $60$\,GeV, 
see\,\cite{AbelleiraFernandez:2012cc}. Two suitable 
configurations have been considered
in the design report: a storage ring mounted on top of the LHC magnets,
the ring-ring configuration (RR), and a separate linac, the 
linac-ring configuration (LR).
In the RR case, bypasses of $1.3$\,km length each are considered
around the existing LHC experiments, also  housing the RF.
This option is now treated as backup
only, mainly because of its strong interference with the LHC. For the LR case,
based on available cavity technology and accepting a synchrotron 
energy loss of about $1$\,\% in the arcs, a new tunnel of racetrack shape and
a length of $9$\,km is required, not much larger than
HERA or the SPS at CERN, see Fig.\,\ref{fig:liview}.
The  tunnel is arranged tangential to IP2 (see below)
and is best positioned
inside the LHC, which avoids a
clash with the LHC injection line TI2 and allows access 
shafts at the Prevessin and Meyrin sites
of CERN, or in close proximity, to be erected. The civil engineering (CE) concepts
were evaluated externally and no principal problem has been observed
which would prevent completion of a tunnel within a few years time.
For the project to begin in the early twenties,
the CE efforts are considered to be strengthened by 2013/14.
\subsubsection*{Components}
Designs of the magnets, RF, cryogenic and further components 
have been considered in some detail. Some major parameters for both the
RR and the LR configurations are summarised in Tab.\,\ref{tabcomp}.
The total number of magnets (dipoles and quadrupoles excluding
the few special IR magnets) and 
cavities is $4160$ for the ring and $5978$ for the linac case.
The majority are the $3080~(3504)$ normal conducting
dipole magnets of $5.4~(4)$\,m length for the ring (linac return arcs),
for which short model prototypes have already been successfully built,
testing different magnet concepts, at
BINP Novosibirsk and at CERN. 
The number of high quality cavities for the two linacs is $960$,
grouped in $120$ cavity-cryo modules.
The cavities of $1.04$\,m length are operated at a currently preferred
frequency of $721$\,MHz, at a gradient of about $20$\,MV/m
in CW mode, as is required for  energy recovery.
The cryogenics system of the ring accelerator is of modest demand.
For the linac it critically depends on the cooling power per cavity, which for
the draft design is assumed to be $32$\,W at a temperature of $2$\,K.  This
leads to a cryogenics system with a total electric grid power of $21$\,MW.
The projected development of a cavity-cryo module for the LHeC
is directed to achieve a high $Q_0$ value and to reduce the
dissipated heat per cavity, which will reduce the dimension of
the cryogenics system.
\begin{table}[hbt]
   \centering
   \begin{tabular}{|l|c|c|}
       \hline
       &  Ring &  Linac \\ 
       \hline
      magnets  & & \\
\hline
%beam energy & \multicolumn{2}{c}{$60 \ \rm{GeV}$}   \\ 
number of dipoles  &  $ 3080 $ & $ 3504 $   \\ 
dipole field [T] & $0.013-0.076$ &  $0.046-0.264$ \\
number  of quadrupoles  &  $ 968 $ & $ 1514 $   \\ 
%field gradient [T/m] & $ 19 $ &  $linac$ \\
\hline
      RF and cryogenics & & \\
\hline
number of cavities & $112$ &  $960$ \\
gradient [MV/m]  & $11.9$ &  $20$ \\
linac grid power [MW] & $ - $ & $24$ \\ 
synchrotron loss compensation [MW] & $49$ & $23$ \\ 
cavity voltage  [MV]  & $5$ & $20.8$ \\  
%temperature   & \multicolumn{2}{c}{$2 \ \rm{K}$}   \\
cavity $R/Q$ [$\Omega$] & 114  & $285$ \\ 
cavity $Q_0$ & $-$ & $2.5~10^{10}$ \\ 
cooling power [kW]  & $5.4$@$4.2$ K & $30$@$2$ K  \\ 
\hline
   \end{tabular}
   \caption{
\footnotesize{
Selected components and parameters of the electron accelerators
 for the $60$\,GeV $e$ beam energy.}
}
   \label{tabcomp}
\end{table}
\subsubsection*{Interaction Region and Choice of IP}
Special attention is devoted to the interaction region design, which
comprises beam bending, direct and secondary synchrotron radiation,
vacuum and beam pipe demands. Detailed simulations are
presented in \cite{AbelleiraFernandez:2012cc}
 of synchrotron radiation effects, which will
have to be pursued further.  Stress simulations, geometry and
material development considerations are presented for
the detector beam pipe, which in the LR case is very asymmetric
in order to accommodate the synchrotron radiation fan.
The LR configuration requires a
long dipole, currently of $\pm 9$\,m length in both directions
from the interaction point, to achieve head-on $ep$ collisions.
The dipole has been integrated in the LR detector concept.
The IR  requires a number of focusing
magnets with apertures for the two proton beams and field-free
regions through which to pass the electron beam.
The field requirements for the RR option 
(gradient of $127$\,T/m, beam stay-clear of $13$\,mm 
($12$\,$\sigma$), aperture radius of $21~(30)$\,mm for the 
$p~(e)$ beam) allow a number of different magnet designs 
using proven $NbTi$ superconductor technology 
and make use of  cable ($MQY$) developments for the LHC.
The requirements for the linac are more demanding in terms
of field gradient (approximately twice as large) 
and tighter aperture constraints
which may be better realised with $Nb_3Sn$ superconductor technology,
requiring prototyping. 

The detector requires an interaction point for $ep$ collisions
while the LHC runs. 
There are eight  points with adjacent long straight tunnel
sections, called IP1-IP8,
that could in principle be used for an experimental apparatus.
Four of these (IP1, IP2, IP5 and IP8)
house the current LHC experiments.
There is no experimental cavern at IP3 nor IP7, 
and it is not feasible to consider
excavating a new cavern while the LHC operates. Since IP6
houses the beam extraction (dumps) and IP4 is occupied with RF
equipment, the LHeC project can only be realised according to  the
present understanding if it uses one of the current experimental
halls. The nature of the $ep$ collider operation is to run
synchronously with $pp$ in the high luminosity phase of the
LHC, which is determined primarily by the searches for ultra-rare phenomena
by ATLAS (IP1) and CMS (IP5). A $9$\,km tunnel excavation and 
surface installations 
close to an international  airport, as would be required at IP8, 
is considered not to be feasible. Therefore, IP2 has been used as the
reference site for the CDR. IP2 appears to be well suited as it has
an experimental surface hall for detector pre-assembly and
with the LHeC inside the LHC ring, access to the linacs
seems to be possible with shafts and surface installations
placed on, or very close to existing CERN territory.
It therefore has to be tentatively recognised that IP2 is in practice
the only option for housing the LHeC detector. This would
require a transition from the ALICE to the LHeC detector, for which
consultations between ALICE and LHeC have recently been initiated. 
%concluding the ALICE experiment in due time.
%\footnote{
%Consultations have begun between ALICE and LHeC for
%discussing the vision of a consecutive $AA$ and $eA$
%physics and transition in the mid twenties.
%}.
%Tentative considerations, as have been noted 
%by NuPECC to this meeting, suggest an end of the ALICE operation
%around the LS3 shutdown time,
%which, with coordinated planning of LS3, of the ALICE and LHeC
%schedules and possibly some dedicated $AA$ LHC operation, 
%makes the conclusion
%of the ALICE programme and subsequent commencement of the LHeC
%during the HL-LHC phase a fruitful vision for the future.}.

The LHeC design report considers only one detector.  This could possibly
be built by two analysis collaborations, cooperating in its 
operation but otherwise ensuring
independent and competing software and
analysis approaches, as
a ``push-pull'' detector philosophy is not feasible.
\subsection{Detector}
\begin{figure}[th]
\begin{center}
\includegraphics[width=.95\columnwidth]{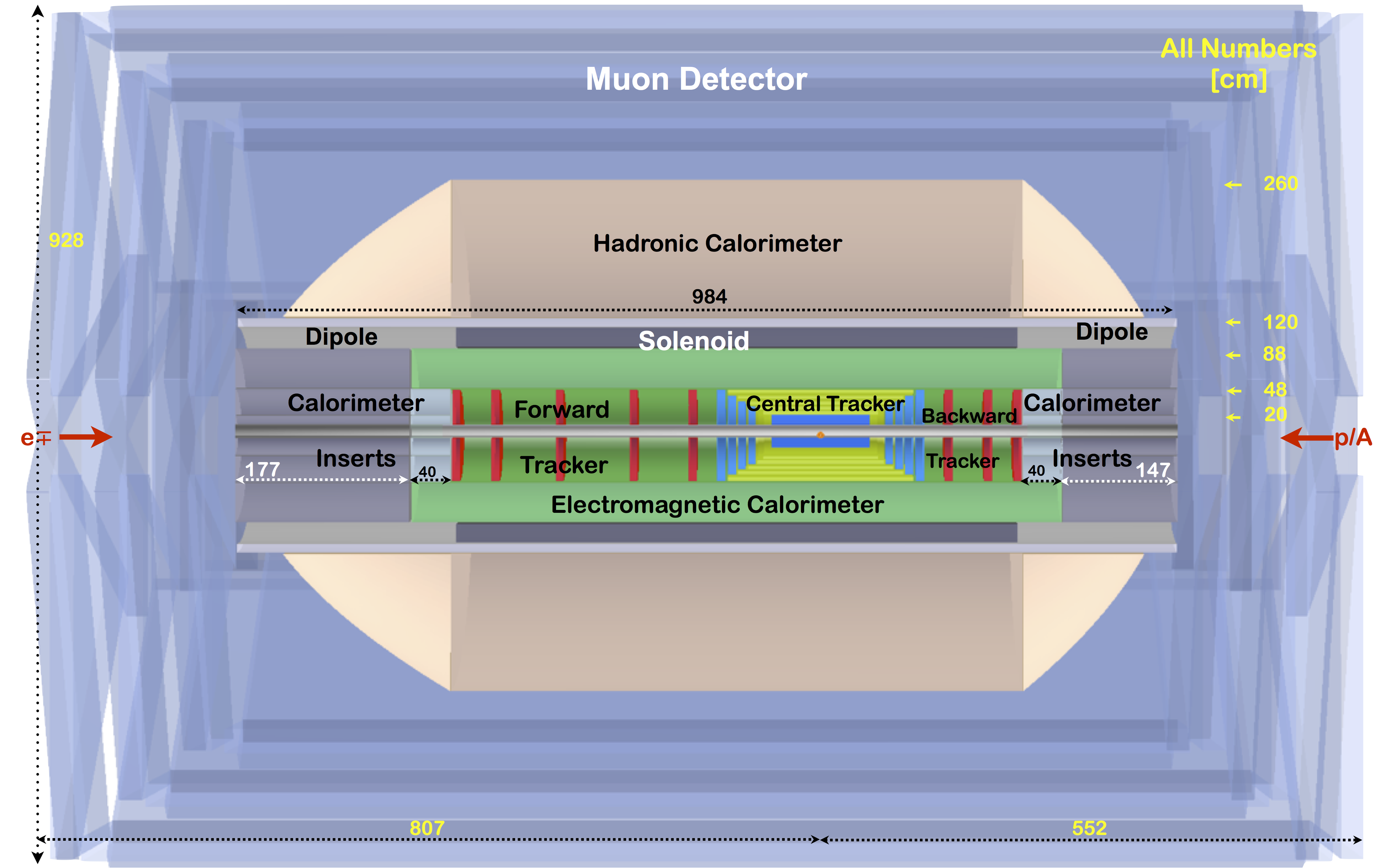}
\end{center}
\caption{
\footnotesize{
An $rz$ cross section of  the LHeC detector
in its baseline design with the magnet configuration for LR, with
the solenoid and dipoles  placed between the electromagnetic and the hadronic
calorimeters. The proton beam, from the right,
collides with the electron beam, from the left, at the IP which is surrounded by
a central tracker system, complemented by large forward and backward tracker
telescopes, and followed by sets of calorimeters.
The detector dimensions are $\approx13.6$\,m in $z$ and a diameter of $\approx9.3$\,m,
which fits in the L3 magnet structure considered for supporting the LHeC
apparatus~\cite{AbelleiraFernandez:2012cc}.
}}
\label{figdet}
\end{figure}

The physics programme depends on a high level of precision,
required for example for the measurement of  $\alpha_s$, 
and on the reconstruction of complex final states, 
as appear in  charged current
single top events or in Higgs production and decay into $b$ final states.
The detector acceptance has to extend as close as possible to
the beam axis because of the interest in the physics at
small and large Bjorken $x$.  The dimensions of the
detector are constrained by the radial extension
of the beam pipe, in combination with maximum
polar angle coverage, down 
to about $1^{\circ}$ and $179^{\circ}$ for forward going final
state particles and backward scattered electrons at low $Q^2$,
respectively. A cross section of the central, baseline detector
is given in Fig.\,\ref{figdet}.
 In the central barrel, the following
detector components are currently considered:
a central silicon pixel detector surrounded by
silicon tracking detectors of strip or possibly strixel technology;
an electromagnetic LAr calorimeter inside
 a $3.5$\,T solenoid and
a dipole magnet required to achieve head-on collisions; 
a hadronic tile calorimeter  serving also 
for the solenoid flux return;
a muon detector, so far only used for muon identification,
relying on the precise inner tracking for muon momentum measurements.
The electron at low $Q^2$ is scattered into the
backward silicon tracker  and its energy is measured
in backward calorimeters. In the forward region,
components are placed for tracking and for
calorimetry to precisely reconstruct jets 
over a wide energy range up to O(TeV).
Simulations of tracking and calorimeter performance
are used to verify the design, although a complete
simulation is not yet available. The report also contains designs
for forward and backward tagging devices for diffractive
and neutron physics and for photoproduction and
luminosity determinations, respectively.
The time schedule of the LHeC project 
demands a detector to be ready within about ten years.
The radiation level at the LHeC is lower than in $pp$,
%, less than $10^{14}$\,n/cm$^2$ equivalent, 
and the $ep$ cross section is low enough for the experiment
not to suffer from
pile-up, which are the two most demanding constraints
for the ATLAS and CMS 
detector upgrades for the HL-LHC. The choice of components
for the LHeC detector  can  rely on the
experience obtained at HERA, at the LHC, including 
its detector upgrades currently being developed, and also on 
detector development studies for the ILC. 
The detector development, while requiring prototyping,
may yet proceed without an extended R\&D program.

A first study has been made about the principles of pre-mounting
the detector  at the surface, lowering  and installing it at IP2.
The detector is small enough to fit into the L3 magnet 
structure of $11.2$\,m diameter, which is still
resident in IP2 and is available as mechanical support.
Based on the design, as detailed in the CDR, it is estimated
that the whole installation can be done in $30$\,months, 
which is compliant with the operations currently foreseen
in the LS3 shutdown, during which 
ATLAS intends to replace its complete inner tracking system.

\subsubsection*{Time Schedule and Mode of Operation}  
The electron accelerator and new detector require a period of
about a decade to be realised, based on experience from previous
particle physics experiments. 
This duration fits with the industrialisation and
production schedules, mainly determined by  
the required $\sim 3500$ approximately $5$\,m long warm
arc dipoles and  the $960$ cavities for the Linac.
The current lifetime estimates for the LHC predict two more decades of operation.
An integrated luminosity for the LHeC of O($100$)\,fb$^{-1}$
may be collected in about one decade. This 
and the current shutdown planning of the LHC define the basic
time schedule for the LHeC project: it has to be installed
during the long shutdown LS3 of the LHC, currently scheduled
for ~2022 and a period of about $2$ years. The connection of the
electron and proton beams and the detector installation can be
realised in a period not significantly exceeding this tentative
time window.
The considerations of beam-beam tune shifts show that the $ep$ operation
may proceed synchronously with $pp$. Therefore with the electron      
beam, the LHC will be turned into a three beam facility.
In the design considerations~\cite{AbelleiraFernandez:2012cc} 
it has been excluded to operate
$ep$ after the $pp$ programme is finished, a) 
because this would make
the LHeC as part of the LHC much
more expensive by adding an extra decade
of LHC operation  requiring also substantial
efforts to first consolidate the LHC, when the high radiation 
$pp$ programme is over, and b) since one would loose 
the intimate and possibly crucial connection
between the $ep/pp$ and the $eA/AA$ physics
programmes, as sketched in this note.

\subsection{Synergies}
The LHeC represents a natural extension to the 
LHC, offering  maximum exploitation of the existing LHC 
infrastructure at CERN. This is a unique advantage as compared
to when HERA was built, for example. 
Physics-wise it is  part of the exploration of the
high energy frontier and as such linked to the LHC and the
lepton-lepton colliders under consideration, a relation
which resembles the intimate connection of HERA to the
physics at Tevatron and LEP for the investigation of
physics at the Fermi scale. 
As an $ep$ and $eA$ machine,
the LHeC unites parts of the particle and  nuclear
physics communities for a common big project. 
It has a characteristic
electroweak, QCD and nucleon structure physics programme which
is related primarily to the LHC but
also to lower energy fixed target DIS experiments
operating at CERN and Jlab, and also to plans for realising
lower energy electron-ion colliders at BNL and at Jlab.
The superconducting (SC) IR magnets resemble HL-LHC  superconducting
magnet developments by the USLARP and SC magnet developments elsewhere.  
The LHeC linac is relevant 
to a variety of projects such as  the XFEL at DESY, ESS, the CEBAF upgrade
at Jlab, the SPL at CERN and other projects for
high quality cavity developments. Through the development of
its  high energy ERL application to particle physics,
the LHeC is related to  about ten lower energy
projects worldwide, which are developing the energy 
recovery  concept. The detector technology is linked mainly to the
LHC experiments and some of their upgrades.
It is thus evident that there are very good prospects
for realising the LHeC within dedicated international
collaborations at a global scale where mutual benefits
can be expected at many levels. The dimension of the LHeC and the
technologies involved make it a suitable project for particle
physics to develop and expand its collaboration with industry.

\section{Physics Highlights of the LHeC}
\subsection{Summary of the Physics Programme}
The LHeC represents a new laboratory for exploring a hugely
extended region of phase space 
with an unprecedented high luminosity
in high energy DIS. 
It is the link between the LHC and a future lepton collider
fulfilling the role played by HERA with the Tevatron and LEP,
but with much higher precision in an extended kinematic range.
%
%It builds the link to the
%LHC and a future pure lepton collider, similar to 
%the complementarity between
%HERA and the Tevatron and LEP, yet
%with much higher precision in an extended energy range. 
%
Its physics is
fundamentally new, and it is also complementary
especially to the LHC, for which the electron beam is an upgrade.
It addresses a broad range of physics questions and an
attempt for a schematic overview on the LHeC physics programme as seen from today's perspective
is presented in Tab.\,\ref{tabphys}. The conquest of new regions of phase space
and intensity has often lead to surprises, which tend to be difficult to tabulate.
%
% -------
%
\begin{table}[hbt]
   \centering
   \begin{tabular}{|l|l|}
       \hline
QCD Discoveries &  $\alpha_s < 0.12$, $q_{sea} \neq \overline{q}$, instanton, odderon, low $x$: (n0) saturation, $\overline{u} \neq \overline{d}$  \\
Higgs & $WW$ and $ZZ$ production, $H \rightarrow b \overline{b}$, $H \rightarrow 4l$, CP eigenstate \\ 
Substructure & electromagnetic quark radius, $e^*$, $\nu^*$, $W$?, $Z$?, top?, $H$? \\
New and BSM Physics & leptoquarks, RPV SUSY, Higgs CP, contact interactions, GUT through $\alpha_s$ \\
Top Quark & top PDF, $xt = x\overline{t}$?, single top in DIS, anomalous top \\
\hline
Relations to LHC & SUSY, high $x$ partons and high mass SUSY, Higgs, LQs, QCD, precision PDFs  \\
\hline
Gluon Distribution  & saturation, $x \eqsim 1$, $J/\psi$, $\Upsilon$, Pomeron, local spots?, $F_L$, $F_2^c$ \\
Precision DIS & $\delta \alpha_s \simeq 0.1$\,\%, $\delta M_c \simeq 3$\,MeV, $v_{u,d},~a_{u,d}$ to $2-3$\,\%, $\sin^2\Theta(\mu)$, $F_L$, $F_2^b$ \\
\hline
Parton Structure & Proton, Deuteron, Neutron, Ions, Photon \\
Quark Distributions & valence $10^{-4} \lesssim x \lesssim 1$, light sea, $d/u$, $s = \overline{s}$\,?, charm, beauty, top \\
QCD & N$^3$LO, factorisation, resummation, emission, AdS/CFT, BFKL evolution \\
\hline
Deuteron & singlet evolution, light sea, hidden colour, neutron, diffraction-shadowing \\
Heavy Ions & initial QGP, nPDFs, hadronisation inside media, black limit, saturation \\
Modified Partons & PDFs ``independent" of fits, unintegrated, generalised, photonic, diffractive \\
\hline
HERA continuation & $F_L$, $xF_3$, $F_2^{\gamma Z}$, high $x$ partons, $\alpha_s$, nuclear structure, .. \\
       \hline
   \end{tabular}
   \caption{
\footnotesize{
Schematic overview on key physics topics for
investigation with the LHeC.}
}
   \label{tabphys}
\end{table}

\subsection{Parton Distributions and the Strong Coupling Constant}
\label{secmax}
\begin{figure}[t]
\vspace{-1.5cm}
\includegraphics[clip=,width=1.\textwidth]{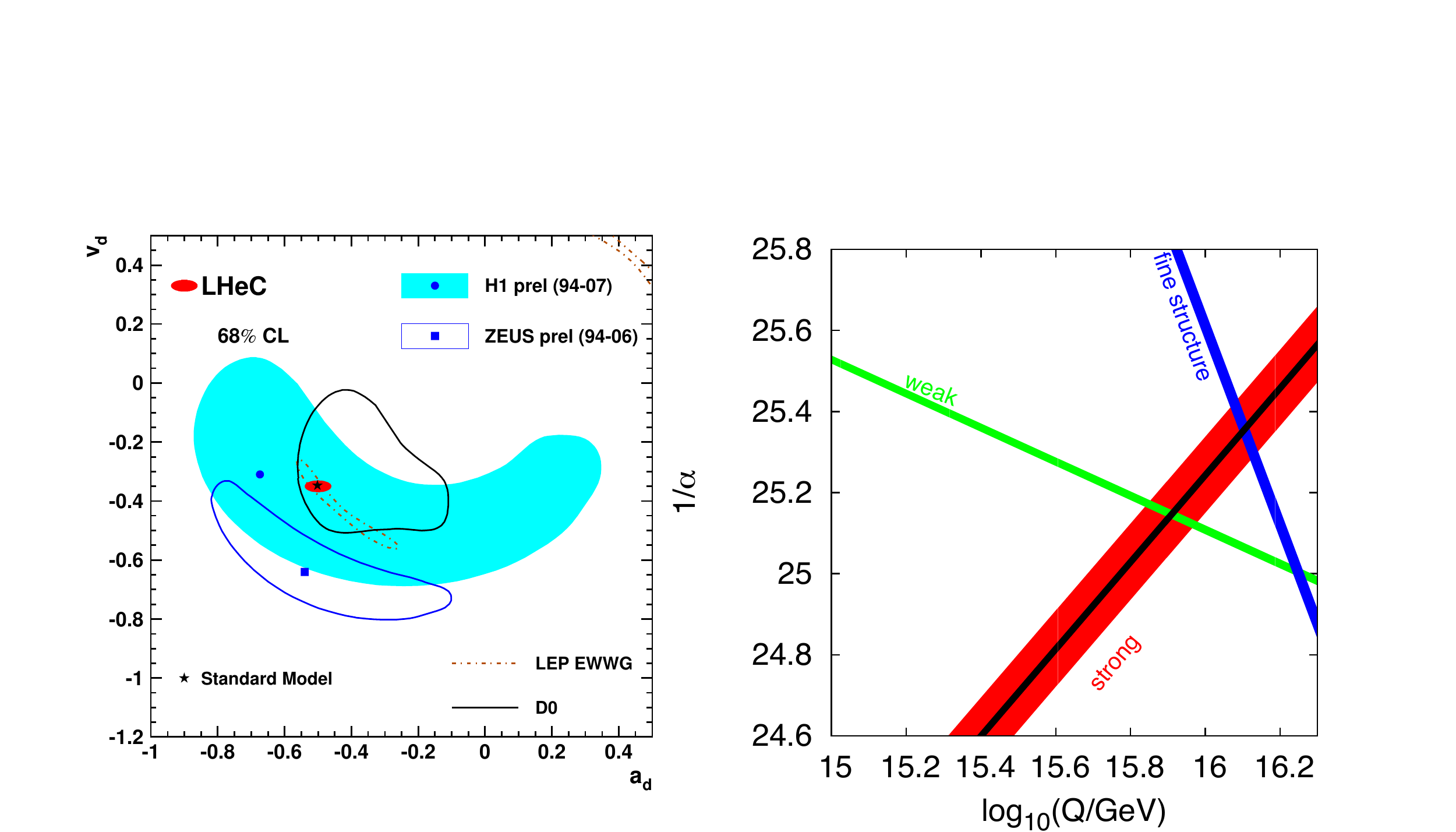}
\caption{
\footnotesize{
Precision electroweak and strong interaction
coupling determinations with the LHeC.
Left: Total experimental uncertainty of the vector and axial-vector
NC down-quark couplings from the LHeC (red ellipse)
compared to present determinations from HERA, Tevatron and LEP;  
% of the hitherto unknown 
% strange-quark distribution;
  Right: Extrapolation of the coupling constants (1/$\alpha$) within 
 SUSY (CMSSM40.2.5)~\cite{AbdusSalam:2011fc}
 to the Planck scale. The width of the red line is the
 uncertainty of the world average of $\alpha_s$, which
 is dominated by
 the lattice QCD calculation chosen for the PDG average.
 The black band is the LHeC projected experimental 
uncertainty~\cite{AbelleiraFernandez:2012cc}.
}}
\label{figstr}
\end{figure}

Despite a series of deep inelastic scattering experiments with 
neutrinos, electrons and muons using stationary targets
and with HERA, and despite the addition of some
Drell-Yan data, the knowledge of the various quark distributions
in the proton is still limited. Due to the wide
kinematic range, huge luminosity, and possibility of beam variations,
the LHeC will provide the necessary constraints on
all parton (quark and gluon) distributions to determine PDFs
completely, free of conventional 
QCD fit assumptions, which has hitherto not been possible.
For example, the valence quarks can be measured up to low and high  $x$,
and the heavy quark distributions, $xs,~xc,~xb$ and $xt$,
 can be  determined from dedicated
$c$ and $b$ tagging analyses with unprecedented precision,
all for the first time.
%An example, for the strange-quark distribution, is presented in
%Fig.\,\ref{figstr}, left.
The QCD fits, which will necessarily evolve with  
real LHeC data, will correspondingly be set-up with a
massively improved and better constrained,
flavour separated input data base, including precision 
electroweak effects, cf. Fig.\,\ref{figstr}\,(left). 
This will have a direct impact on the extension of the 
search range at the LHC at high masses, above a few TeV,
see Sect.\,\ref{secmon}.

From a full simulation of experimental systematic uncertainties,
as is described in~\cite{AbelleiraFernandez:2012cc}, 
one concludes that the strong coupling
constant $\alpha_s(M_Z^2)$ can be measured to per mille precision, as compared
to a per cent level nowadays. Such a measurement puts the attempts
to unite the couplings at the Planck scale on a firm footing for the first
time, see Fig.\,\ref{figstr}\,(right). Combined with a similarly
high accuracy from jet data, this measurement will eventually resolve
the question of whether inclusive DIS and jet-based $\alpha_s$
measurements lead to the same or to different strong coupling constants.
It also finally makes the BCDMS measurement redundant, which
for a long time pointed to a very small value of $\alpha_s$
obtained in the most precise measurement in DIS. The LHeC
$\alpha_s$ measurement
is not just a single experiment but represents a whole programme,
which renews the  physics of DIS and revisits the scale
uncertainties in pQCD at the next-to-next-to-next-to leading
order level. The LHeC itself provides the necessary basis
for such a programme, mainly with a complete 
set of high precision PDF measurements, including
for example the prospect to measure the charm mass
to $3$\,MeV as compared to $30$\,MeV at HERA (from $F_2^{cc}$),
and with the identification of the limits of applicability of DGLAP
QCD by discovering or rejecting saturation of the gluon density.

\subsection{Low \bf{$x$} Physics}
\label{secpaul}
The parton densities extracted from HERA data
exhibit a strong rise towards  
low $x$ at fixed $Q^2$. The low $x$ regime of 
proton structure is a largely
unexplored territory whose dynamics are those of a densely 
packed, gluon dominated, partonic system.  
It offers unique insights into the gluon field which
confines quarks within hadrons and 
is responsible for the generation of most of the mass of hadrons. 
Understanding low $x$ proton structure is also important 
for the precision study of cosmic ray air showers and ultra-high
energy neutrinos and may be related to the string theory of gravity.
The most pressing issue in low $x$ physics is the need for a mechanism
to tame the growth of the partons, which,
from very general considerations, 
is expected to be modified in  
the region of LHeC sensitivity. 
There is a wide, though non-universal, consensus, that
non-linear contributions
to parton evolution (for example via gluon recombinations $gg \rightarrow g$) 
eventually become relevant and the parton densities `saturate'. 
The LHeC offers the unique possibility of
observing these non-perturbative dynamics at sufficiently large 
$Q^2$ values for weak coupling theoretical
methods to be applied, suggesting the
exciting possibility of a 
parton-level understanding of the collective properties of QCD. 
A two-pronged approach to mapping out the newly accessed LHeC low
$x$ region is proposed in \cite{AbelleiraFernandez:2012cc}. On the one
hand, the density of partons can be increased by overlapping
many nucleons in $eA$ scattering (see next section). On the other
hand, the density of a single 
nucleon source can be increased by probing at
lower $x$ in $ep$ scattering. Many observables are considered in
\cite{AbelleiraFernandez:2012cc}, from which two illustrative examples
are chosen here. 
\begin{figure}[h]
\includegraphics[clip=,width=.47\textwidth]{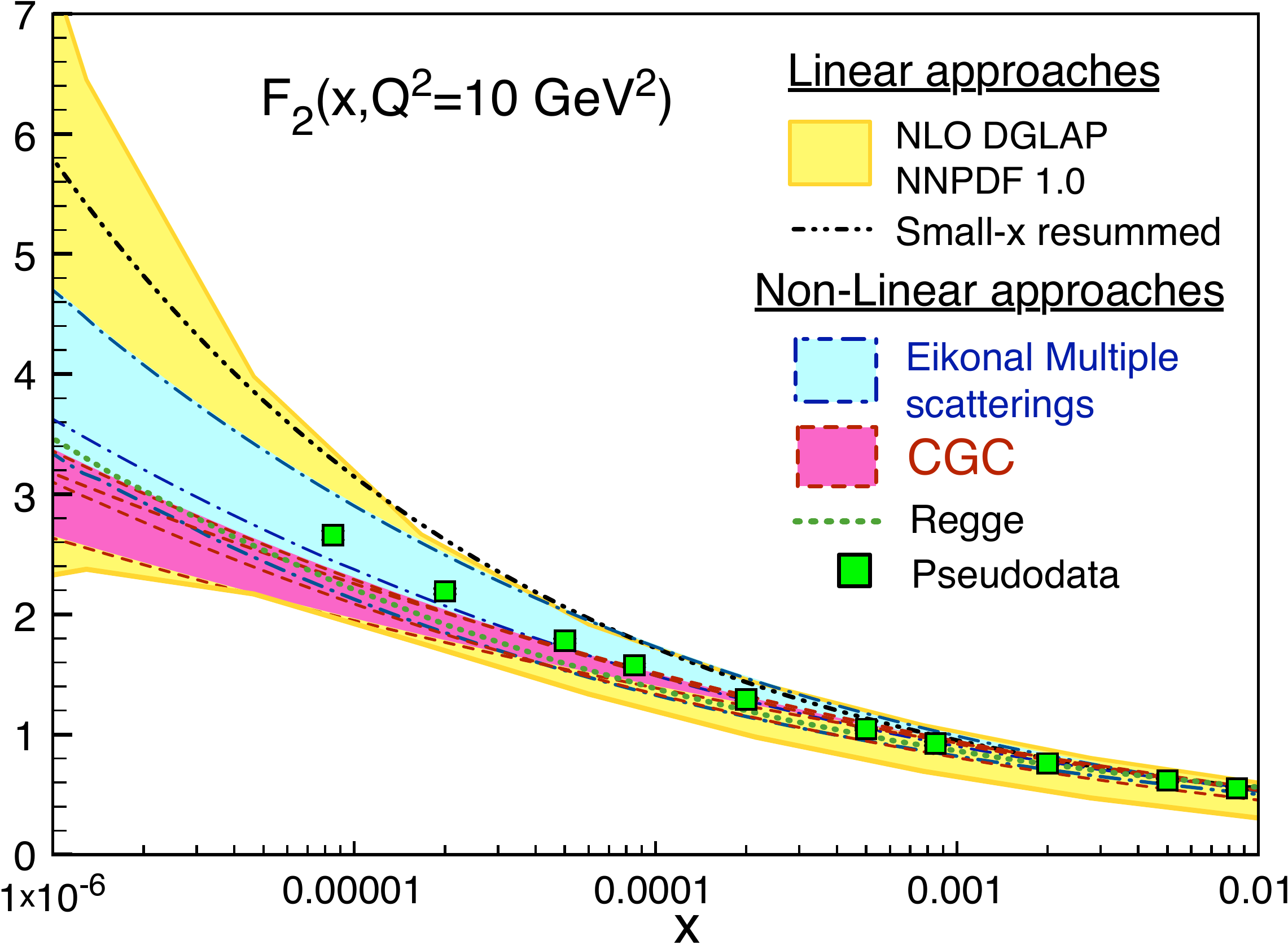}
\includegraphics[clip=,width=.55\textwidth]{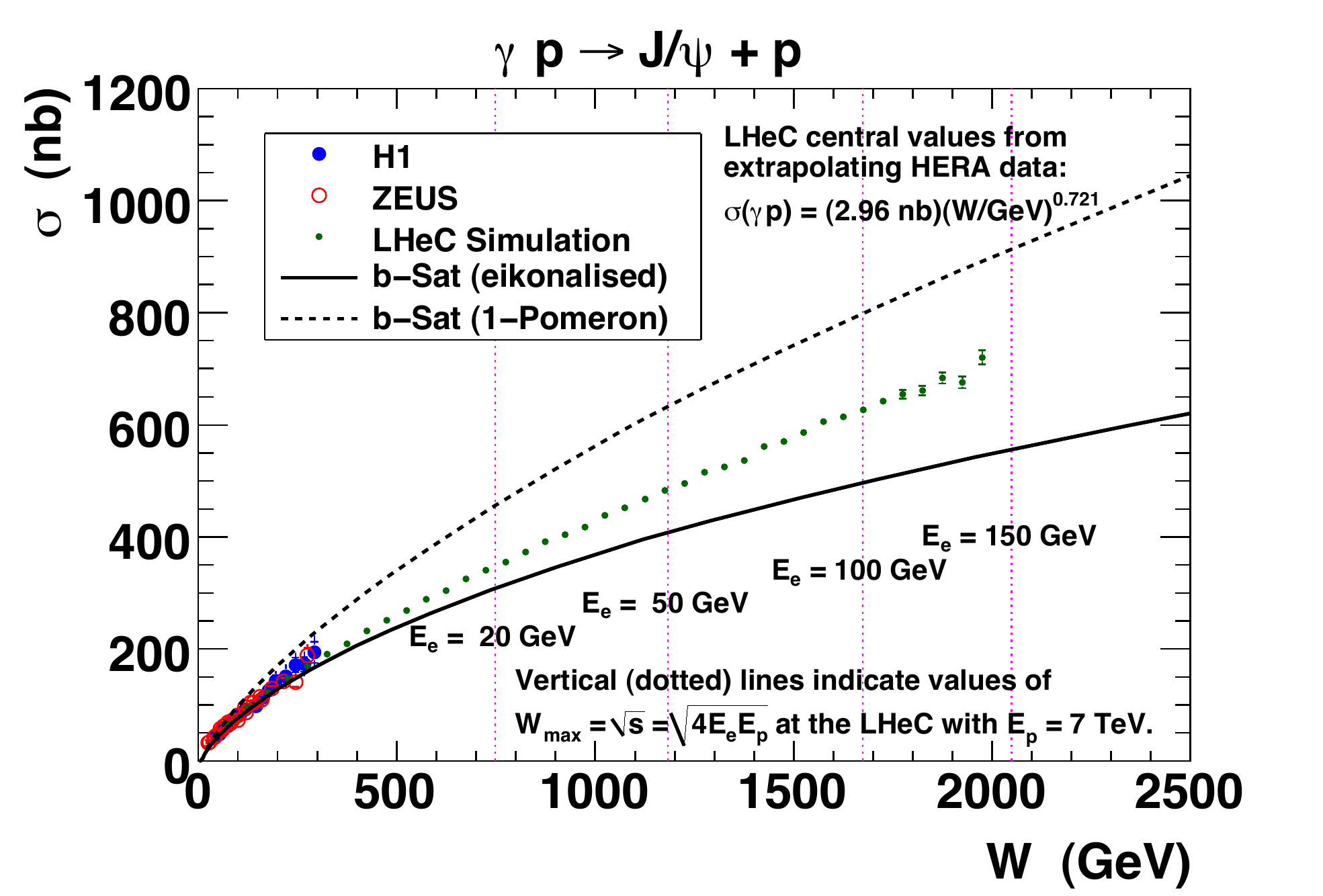}
\caption{
\footnotesize{
Left: Simulated low $x$ LHeC $F_2$
data at a single illustrative value of $Q^2$, compared with predictions
from standard 'linear' DGLAP-based QCD and from models 
which include saturation
effects. Right: Simulated LHeC measurements of the 
exclusive $J/\psi$ photoproduction cross section, compared with models 
which include or neglect saturation phenomena.}
}
\label{lowx:figs}
\end{figure}

Fig.~\ref{lowx:figs} (left) shows simulated low $x$ LHeC $F_2$ data 
at a single example
value of $Q^2$, compared with three bands of predictions.
The `NNPDF 1.0' band is one estimate of the range of variations possible
from low $x$ extrapolations of current NLO DGLAP QCD fits. 
The `Eikonal Multiple scatterings' and `CGC'
bands are envelopes spanning numerous predictions based on two 
different QCD-based approaches to the onset of non-linear
dynamics and parton saturation. It is clear that the LHeC data are
more than adequate to distinguish between these different models, 
particularly when complementary precision
data on the $F_L$ structure function
(not shown here, see \cite{AbelleiraFernandez:2012cc}) are also included. 
Fig.~\ref{lowx:figs} (right) shows simulated LHeC measurements of the cross
section for the exclusive photoproduction of $J/\psi$ mesons
($\gamma p \rightarrow J/\psi p$, 
obtained as the $Q^2 \rightarrow 0$ limit of the process
$e p \rightarrow e J/\psi p$), as a function of the $\gamma p$ 
centre of mass energy, $W$. The `b-Sat (1-Pomeron)' and
`b-Sat (eikonalised)' curves correspond to predictions constrained
by HERA data which either neglect or include saturation effects, 
respectively. The conclusion is that LHeC data will not
only unambiguously establish the onset of saturation phenomena, but
will be highly sensitive to the, presumably rich, underlying dynamics.
Moreover, the extraction of generalised Parton Distributions (GPDs) - for
both quarks and gluons - accessible in hard exclusive reactions such as deeply virtual
Compton scattering and meson electroproduction will allow to understand proton structure in a
new, three-dimensional way.

\subsection{Electron-Ion Scattering}
The LHeC will give access to a completely new kinematic regime in the $x-Q^2$ plane, 
compared to previous  experimental facilities, for
exploration of nuclear structure and dynamics of nuclear collisions. 
As illustrated in Fig.~\ref{Fig:eAkinplane}\,(left),
the gain towards smaller $x$ and larger $Q^2$ is almost four orders of magnitude.
Besides, the $eA$ mode offers the possibility of a two-pronged approach to explore the
small-$x$ dynamics of QCD, where a novel regime, characterised by the
 breakdown of linear fixed-order perturbation theory applicable in a 
dilute parton region, and the appearance of non-linear evolution phenomena 
and saturation of partonic densities, is expected to exist. With all these 
features being triggered by an increase of parton densities, at the LHeC this novel regime 
can be approached, as shown in Fig.~\ref{Fig:eAkinplane}\,(right), through
a decrease in $x$ and through an increase in the number of nucleons, $A$, involved. 
The latter can be achieved in $eA$ collisions both increasing $A$ and decreasing 
the impact parameter of the collision - thus the effective number of nucleons 
involved. For this reason, the physics case presented in~\cite{AbelleiraFernandez:2012cc} 
for $ep$ collisions has been made in parallel for $eA$ whenever possible. 
As shown in~\cite{AbelleiraFernandez:2012cc,chavannes12}, the possibilities 
for nuclear studies at the LHeC are vast:
\begin{figure}[h*]
\centerline{\includegraphics[clip=,width=0.40\textwidth,angle=0]{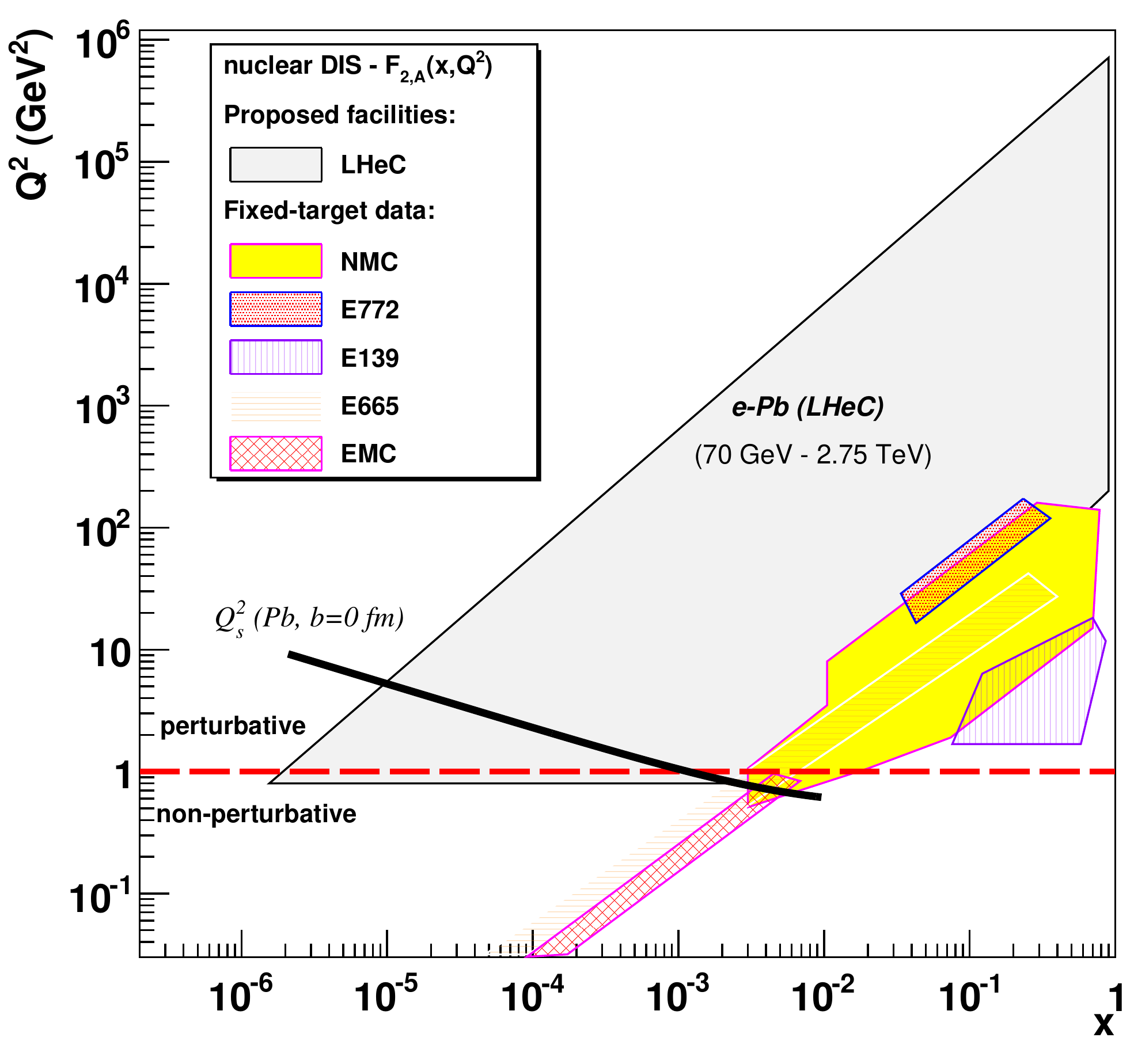}\hskip 1cm\includegraphics[clip=,width=0.45\textwidth,angle=0]{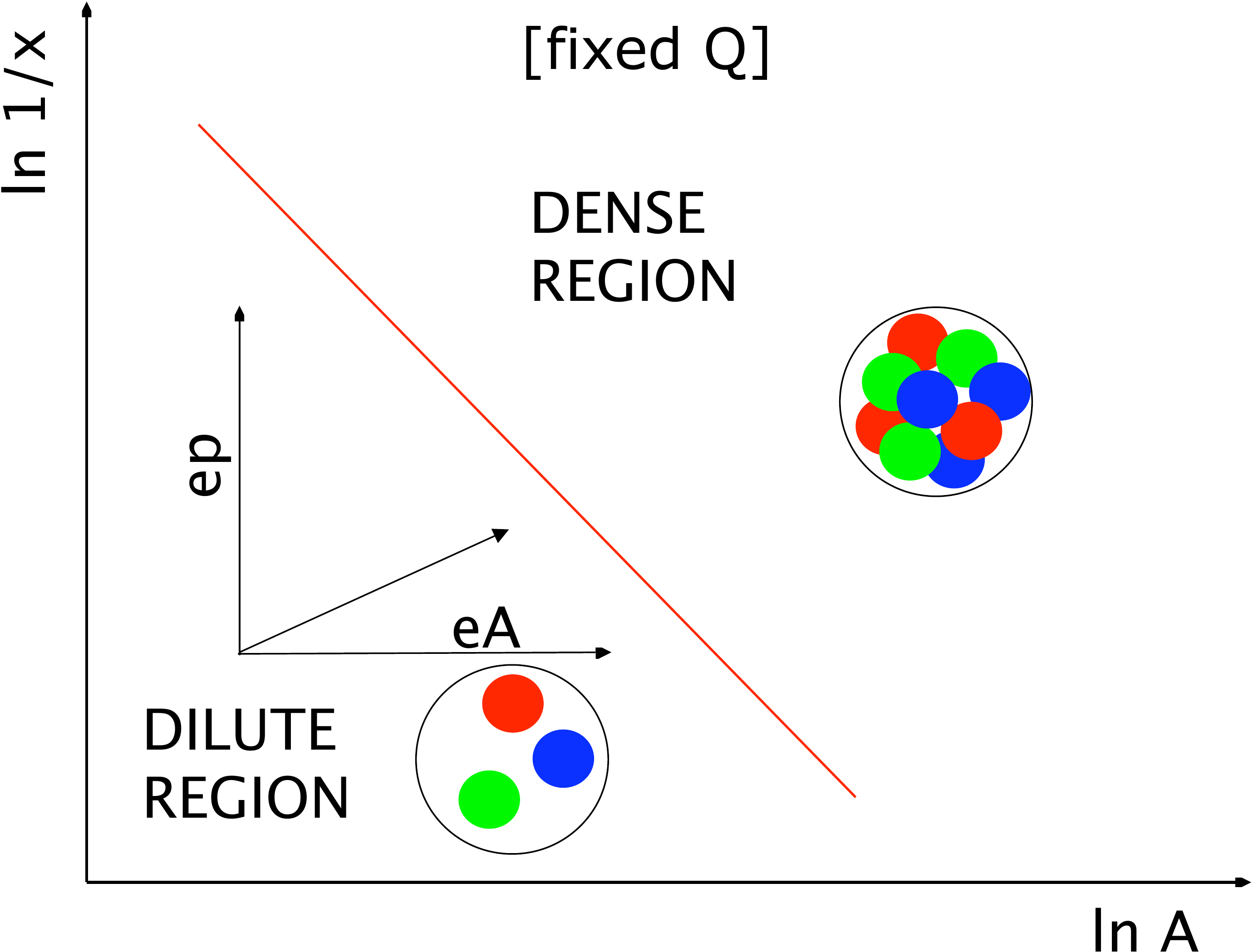}}
\caption{
\footnotesize{
Left: kinematic coverage of the LHeC in the $\ln Q^2 - \ln 1/x$ plane 
for nuclear beams, compared with existing nuclear DIS and Drell-Yan experiments.
The thick black line locates roughly the expected transition region
between a dilute partonic regime above and a dense regime below. Right: 
schematic view of the different regions for the parton densities in 
the $\ln 1/x - \ln A$ plane, for fixed $Q^2$. 
Lines of constant occupancy of the hadron are parallel to the diagonal 
line shown.}}
% Plots taken from~\cite{AbelleiraFernandez:2012cc}.}
\label{Fig:eAkinplane}
\end{figure}

\noindent $\bullet$ Structure function measurements and their flavour decompositions in 
$eA$ will allow  nuclear parton densities at small $x$ to be measured for the
first time, testing current extractions through linear evolution equations,  
particularly for the presently almost unconstrained gluon density for $x<10^{-2}$,
and the unknown charm and beauty densities in nuclei.

\noindent $\bullet$ Exclusive vector meson production in $eA$ will offer a handle,
complementary to $ep$, on the possible evidence of non-linear dynamics and saturation 
of partonic densities, as these effects increase with $A$.

\noindent $\bullet$ Inclusive diffraction in $eA$ will be measured 
for the first time, with the possibility of extracting diffractive 
parton densities in nuclei, and of looking for signals of novel high-density 
phenomena which are expected to affect diffraction to a larger extent than inclusive observables. 
%For both inclusive and exclusive diffraction, the separation of  coherent, with the nucleus remaining intact, from incoherent, with nuclear breakup, will be achieved through a combination of central and forward detectors.

\noindent $\bullet$ The dynamics of hadronisation and QCD radiation will be clarified in
$eA$ collisions through semi-inclusive measurements of both particles and jets, of 
which large yields will be produced up to high transverse momenta. The 
effects of the nuclear environment will be explored through the modification of 
yields, hadrochemistry, jet substructure, etc., compared to equivalent $ep$ measurements.

All these measurements will settle the role of factorisation, either the 
standard collinear one or some other version suitable for high energies 
and dense regimes,  in $eA$ collisions for inclusive, semi-inclusive and diffractive observables.
%Many other aspects that have been addressed in $ep$
%, as generalized parton densities,  possibilities for discriminating linear from non-linear evolution schemes, QCD dynamics through forward measurements,$\dots$,
%will be subject of future work.
Finally, note the strong implications of the investigation of $eA$ collisions at 
the LHeC on ultrarelativistic heavy-ion physics, both by constraining the initial 
state, which is crucial for the quantitative understanding of
the subsequent behaviour of the produced medium, and 
for our understanding of the ability of hard probes to characterise it. 
Some of these aspects will be explored in $pA$ collisions at the LHC in a similar
kinematical region, but they will be studied  at the LHeC in a much
cleaner manner.
%
%Note the strong implications of the study of $eA$ collisions at the LHeC on ultrarelativistic heavy-ion physics: First, the nuclear wave function that establishes the initial condition for particle production in $AA$ collisions. Second, the mechanism of particle production that specifies the initial stage of the collisions prior to an eventual thermalization or isotropization and the subsequent collective evolution. Finally, the understanding of the nuclear modifications of hadronization and QCD radiation and their effects on particle distributions and jet observables, that determine their ability to characterize the medium produced in such collisions. All these aspects will be explored in $pA$ collisions at the LHC in a similar kinematical region, but they will be studied  at the LHeC in a sizably cleaner manner.

\subsection{Higgs Physics with the LHeC}
\label{sec:higgslhec}
In the Standard Model, the breaking
of the electroweak $SU(2)_L \times U(1)_Y $ symmetry gives mass to the
electroweak gauge bosons via the Brout-Englert-Higgs mechanism while the fermions
obtain their mass via Yukawa couplings with a scalar Higgs field. 
With the observation of a Higgs-like boson by the ATLAS and CMS 
collaborations with a mass around 126\,GeV, a new research field 
has opened in particle physics. The measurement of the couplings 
of the newly found boson to the known fundamental particles will 
be a crucial test of the SM and a window of opportunity to establish
physics beyond the SM. 

%
%\begin{figure}[h]
%\includegraphics[clip=,width=.47\textwidth]{higgs60a4.pdf}
%\caption{higgs
%}
%\label{figH}
%\end{figure}
%
At the LHeC, a light Higgs boson could be uniquely produced 
and cleanly reconstructed either 
via $HZZ$ coupling in neutral current (NC) DIS or via $HWW$ coupling in 
charged current (CC) DIS. Those vector boson fusion processes have 
sizeable cross sections, O(100)\,fb for $126$\,GeV mass,
and they can be easily distinguished, which is a unique advantage 
in comparison to  the VBF Higgs production in $pp$ scattering.
The observability of the Higgs boson signal at the LHeC was 
investigated in the CDR~\cite{AbelleiraFernandez:2012cc} initially using 
the dominant production and decay mode, i.e.
the CC reaction
 $e^- p \rightarrow H (\rightarrow b \bar{b}) + \nu + X$,
for the nominal $7$~TeV LHC proton beam  
and electron beam energies of $60$ and  
$150\,$GeV.
% and instantaneous luminosities at the level 
%of $10^{33}$ cm$^{-2}$s$^{-1}$. 
%At the preferred mass, the Higgs 
%would predominantly decay into a $b \bar{b}$ pair with a branching ratio of about $60$\%.  
Signal and multi-jet background processes are generated 
using MadGraph. The generated events are passed through
 a custom-made Pythia program version for fragmentation and 
hadronisation. The event reconstruction is still based
on a parameterised, generic LHC-style detector. 
%The set-up
% allowed important investigations of basic detector effects
% which are a 
Very good tracking and  $b$-tagging capabilities
%, up to $\eta=3$, 
 are crucial for the signal 
identification and enrichment, while a good hadronic 
energy and flow resolution 
%of about $60\%\sqrt{E/GeV }$
is important for the background rejection and the 
tagging of very forward jets.
%, up to $\eta=5$.
Simple and robust cuts are identified and found to reject 
effectively e.g. the dominant single-top background,
as opposed to refined event weighting and network techniques
presently utilised in complex final state (top and Higgs)
physics at the LHC, providing an excellent S/B ratio of 
about~$1$ at the LHeC.
%With an electron beam of 150\,GeV 84.6 events are expected 
%for 10\,fb$^{-1}$ of integrated luminosity with an 
%excellent signal to background ratio of 1.8  using
% the cut-based analysis, a signal to background ratio 
%far superior to that expected for the LHC for this 
%decay channel.
At the default electron beam energy of 
$60$\,GeV, for $80$\,\% $e^-$ polarisation and an 
integrated luminosity of 100 fb$^{-1}$, 
the $Hbb$ coupling is estimated to be measurable
with a statistical precision of about 4\,\%, which is
not far from the theoretical uncertainty
\footnote{
It is worth emphasising that many of these
uncertainties, as the parametric ones discussed 
in~\cite{Denner:2011mq}, are of PDF and QCD nature and
therefore will be very much reduced by the
results anticipated with the realisation of the LHeC
programme. This should contribute significantly
to improving the accuracy of the Higgs 
measurements with the LHC.}.
Typical coupling measurements,
such as $\gamma \gamma$ or $4l$, can be made with about $10$\,\%
precision with the HL-LHC, while the specific $b\overline{b}$
coupling will be particularly difficult to measure due
to high combinatorial backgrounds in $pp$.

It has also been observed, that the LHeC can
specifically explore well the CP structure 
of the $HWW$ coupling by separating it from the $HZZ$ 
coupling and the other signal production mechanisms. Any 
determination of an anomalous $HWW$ vertex will 
thus be free from possible contaminations of these.
A further advantage of the $ep$ collider kinematics stems 
from the ability to disentangle clearly the direction
 of the struck parton and the final state lepton (clear 
definition of forward and backward directions). Compared 
to the $pp$ situation, $ep$ lacks 
the complications due to underlying event and pile-up
driven backgrounds.

The few initial studies performed so far will be pursued further
in the light of recent observations from the LHC experiments.
%that the Higgs boson is now likely to indeed exist.
For the projected analyses, this primarily concerns using a full
LHeC detector simulation, and optimising further
the detector design. For the accelerator design it is obvious that
a luminosity in excess of   $10^{33}$~cm$^{-2}$s$^{-1}$
is very desirable. 
This would open up the possibility of also making precision measurements
of rarer ($\tau$, $Z$, $W$, perhaps photon) 
decay channels, the CP angular distributions for
both the $HWW$ and $HZZ$ couplings, 
and NC initiated production, 
a scenario in which the LHeC collider would have a truly remarkable potential to study both the Higgs
boson and mechanism.

%An enhancement of the instantaneous luminosity to the 
%level of $10^{34}$ cm$^{-2}$s$^{-1}$ would open an incredible
%  window of opportunity for Higgs physics at the LHeC:
%\begin{itemize}
%\item The accummulation of about a thousand
% $H\rightarrow b\overline{b}$ events would bring down
% the statistical error on the Hbb Yukawa coupling 
%measurement to the level of the theory uncertainty~\cite{Denner:2011mq}. 
%\item The addition of the neutral current 
%$e p \rightarrow H (\rightarrow b \bar{b}) + e + X$ process 
%would further improve the determination of the Hbb Yukawa 
%coupling. This would also lead to the exploration of the CP 
%structure of the HZZ coupling.
%\item The addition of coupling measurements using the decays
% to $\tau$, $Z,W$ and, possibly, $\gamma$ pairs.
%\end{itemize}
%The initial studies performed so far are to be pursued further 
%as the design of the apparatus and its simulation proceed. With
% a luminosity enhancement to the $10^{34}$ cm$^{-2}$s$^{-1}$ 
%level the potential of exploring Higgs physics of the
%LHeC would be striking. 

\subsection{The Search for SUSY and the LHeC}
\label{secmon}
Supersymmetry (SUSY) is a compelling theory providing an 
extension of the Standard Model (SM) at arbitrarily high energies. 
In the SM, the Higgs boson mass suffers from large quantum loop corrections,
as large as the cut-off scale of the theory, requiring a high 
degree of fine-tuning of parameters in order to be cancelled.
% -- the so-called `naturalness problem'. 
In SUSY, the addition of supersymmetric partner particles to the known fermions 
and bosons cancels the largest of these loop effects, permitting the Higgs boson 
mass to lie naturally at the $\sim 100\,$GeV scale. 
Supersymmetric models present other advantages, including improved unification of running 
coupling constants at the Planck scale, see Sect.\,\ref{secmax},
 and renormalisation group equations that radiatively 
generate the scalar potential that leads to electroweak symmetry breaking. 

The possible conservation of $R$-parity, a discrete quantum number which relates 
spin (S), baryon and lepton numbers (B and L), is fundamental in determining the phenomenology of SUSY. 
In the framework of generic $R$-parity conserving supersymmetric extensions 
of the SM, SUSY particles are produced in pairs and the lightest supersymmetric 
particle (LSP) is stable. In a large variety of models the LSP is the 
lightest neutralino, $\tilde{\chi}^{0}_{1}$, one of the SUSY partners of the gauge bosons 
together with its three heavier mass eigenstates ($\tilde{\chi}^{0}_{2,3,4}$)  
and the charginos ($\tilde{\chi}^{\pm}_{1,2}$). The lightest neutralino only 
interacts weakly and provides a  Dark Matter candidate with the appropriate 
relic density to explain the cosmological Dark Matter. 
The possible appearance of $R$-parity-violating couplings, and hence the non-conservation 
of baryon and lepton numbers ($B$ and $L$) in supersymmetric theories, imply an even 
richer phenomenology. Although $R$-parity-violating interactions must be sufficiently 
small, their most dramatic implication is 
the automatic generation of neutrino masses and mixings. The possibility that the results 
of atmospheric and solar neutrino experiments may be explained by neutrino masses and 
mixings originating from $R$-parity-violating interactions has motivated a large number 
of studies and models. 

The discovery (or exclusion) of supersymmetric particles remains a
high priority for the LHC experiments, the LHC being the primary machine to 
search for physics beyond the SM at the TeV scale. The role of the LHeC is to 
complement and possibly resolve the observation of new phenomena.  
At $ep$ colliders, SUSY particles could be produced due to sizeable lepton flavour 
violating terms or, in the framework of $R$-parity conserving models, via 
associated production of selectrons  and first and 
second generation squarks. The latter process 
would present a sizeable cross section only if the sum of selectron and squark masses 
is below or around 1~TeV (thus, for relatively light squarks). 
Current exclusion limits set by the ATLAS and CMS 
experiments on first and second generation squarks are up to 1.5~TeV under the 
assumption that the first two squark families are degenerate. In models where 
this condition is relaxed, windows of discovery relevant for the LHeC 
might still be open at the time of start-up. 

Less stringent constraints exist in the context of $R$-parity violating scenarios. 
Processes of interest for $ep$ colliders include leptoquark-like processes  
and associated production of quark-neutralinos via squark-quark-neutralino couplings, 
where squarks can be off-shell and thus beyond direct LHC exclusion limits. While stringent 
constraints exist for associated production of leptons and quarks possibly 
deriving from squark decay, processes of the kind $eu \rightarrow d\tilde{\chi}^{0}_{1}$ 
and subsequent decays of $\tilde{\chi}^{0}_{1}$ to leptons and quarks might be 
difficult to study at the LHC, due to the overwhelming SM multi-jet background, 
and can be successfully searched for at the LHeC.   
 
\begin{figure}[!h]
\begin{center}
\includegraphics[clip=,width=.50\textwidth]{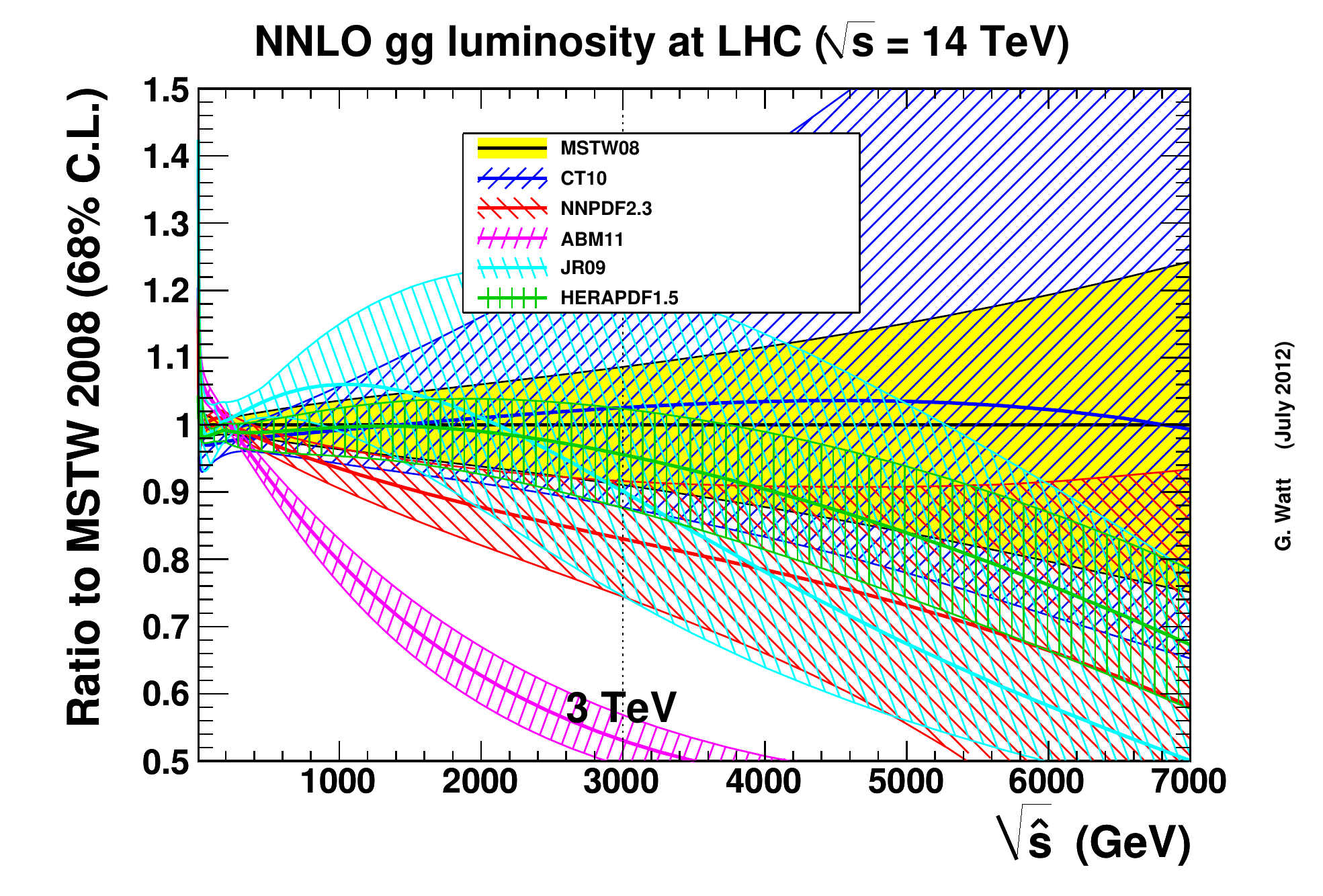}
\includegraphics[clip=,width=.44\textwidth]{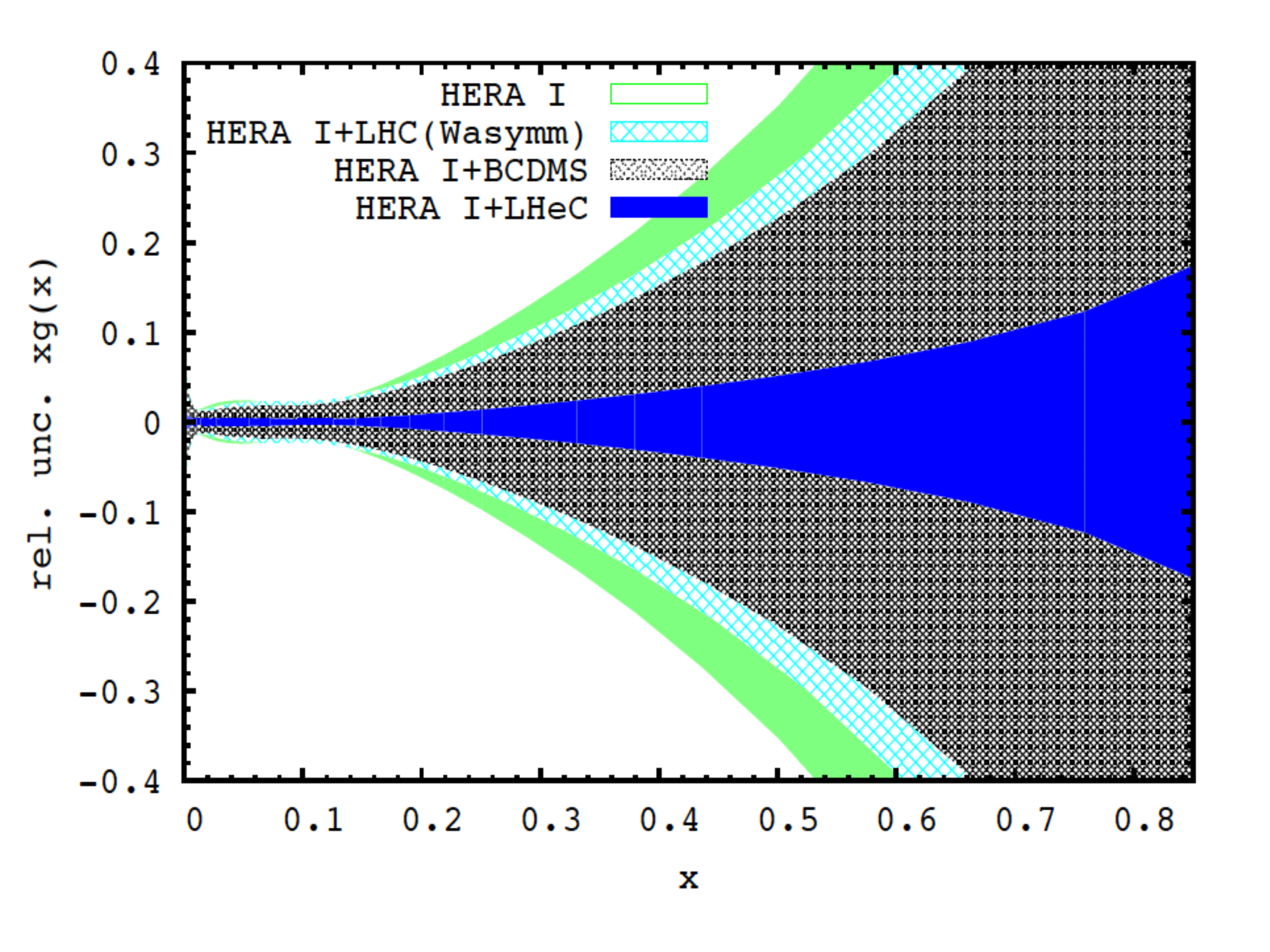} 
\caption{
\footnotesize{Left: Gluon-gluon luminosity calculated as a function of
$\sqrt{\hat{s}}=\sqrt{x_1x_2s}$, where $x_1 x_2$ is the product of the two Bjorken $x$
values with which the two gluons are emitted. The maximum mass one
can pair produce is $M_{\tilde{g}}=\sqrt{x_1x_2s}/2$, i.e. a $4$\,TeV
mass corresponds to an average $x=\sqrt{x_1 x_2}$ of about $0.6$.
 The $gg$ luminosity,
%  defined as
%$L(\hat{s})_{gg} = \int xg(x_1,\hat{s}) xg(x_2,\hat{s}) \delta ( s x_1 x_2 - \hat{s}) dx_1dx_2 $, 
is a measure, here expressed
as the ratio to the MRST08 prediction, of the gluino-pair production
process  $gg \rightarrow g \rightarrow \tilde{g}\tilde{g}$,
which is one of the interesting SUSY channels open for
discovery, with  masses as high as about $4$\,TeV accessible in the HL-LHC phase.
Right: The LHeC will determine the partons at high $x$ very accurately,
together with the strong coupling to per mille precision,
which is demonstrated here with the projected uncertainty
estimated for the gluon distribution at high $x$ from inclusive
DIS alone (right plot), quoted at the starting scale of the evolution.}
}
\label{fig:figxg}
\end{center}
\end{figure}

The LHeC will also provide indirect handles for the case of supersymmetry. 
The dominant SUSY production channels at the LHC are assumed to be 
squark-(anti)squark, 
squark-gluino, and gluino-gluino pair production. All gluon-initiated processes  
suffer from very large uncertainties due to the extremely limited knowledge
of the gluon density at high $x$.
If gluinos of $2-4$\,TeV mass exist, their discovery, e.g. through 
enhanced multi-jet production rates (with or without missing transverse momentum)
compared with SM predictions, and their kinematic characterisation 
will depend on the capability to predict their production cross section with 
good precision. Current PDF fits suffer from very large uncertainties, as shown in 
Fig.\,\ref{fig:figxg} (left), and also differ considerably in their central predictions.
This will be completely changed with the high precision LHeC measurements, which will allow
for the first time assumption-free, all-flavour PDFs, implying improvements in both the agreement between
the range of possible central values as well as the precision.
The improvement in the precision of the
gluon is illustrated in Fig.\,\ref{fig:figxg} (right), while the
improvements for the quarks at high $x$ can be found in \cite{AbelleiraFernandez:2012cc}.
%\newpage
%\noindent
%\vspace{0.5cm}
The interest in $R$-parity violating SUSY translates directly
into the striking potential of the LHeC to determine the
lepto-quark or lepto-gluon quantum numbers should such states
be discovered at the LHC. The $ep$ machine has a clean $s$-channel
single production mode, with variable input beam parameters
while the LHC produces them predominantly in pairs. This
is discussed in detail in \cite{AbelleiraFernandez:2012cc}, 
along with the reach in contact
interactions, excited leptons, anomalous lepton-quark interactions
and other BSM topics.
% in \cite{AbelleiraFernandez:2012cc}. 

%
\newpage
\section{Outlook}
The design report~\cite{AbelleiraFernandez:2012cc}
shows  that the LHeC can be 
realised at CERN and would substantially enrich the
physics accessible with the LHC. 
%Given its connection to the LHC schedule,
%the next few years will be used for
%prototyping critical components, such as the superconducting
%IR magnets, and moving towards an LHeC ERL test facility.
The ongoing development of the most important components 
will facilitate an informed decision 
about the project, expected around 2015 when the
first luminous results from the increased energy LHC
%with $13-14$\,TeV in the cms, 
become available.
In order to then proceed, the whole project
is being further developed,
%correspondingly,
which jointly regards its physics, simulation,
the detector design as well as the interaction
region, civil engineering and many aspects 
%and systems 
of the accelerator. 
% magnets, cavities, the vacuum and beam pipe
%design, civil engineering and further items.

During 2012 discussions
have gained in intensity at CERN and with(in)
the interested community, to evaluate and develop the LHeC
prospect, which is  uniquely linked to the overriding success
and quality of the LHC accelerator. 
Very recently, CERN has issued a
mandate~\cite{sbchavannes}
to the LHeC Study Group 
which is directed to an LHeC
ERL test facility and comprises the
critical components enabling a decision
a few years hence.
Since 2006, the
LHeC has been part of the EU strategy deliberations,
which are now ongoing. The project development has been 
accompanied by
ECFA, the European Committee for Future Accelerators,
and by NuPECC, the Nuclear Physics European Collaboration
Committee, which in 2010 decided to include the
LHeC in its long range plan.

As a new TeV energy scale collider, the LHeC
naturally has all the characteristics of a global project of interest
and importance extending beyond Europe. Its physics
will substantially enrich the LHC programme,
to which it represents an upgrade, and
its technology is fascinating in many respects.
With the addition of an electron beam, CERN
would provide the world particle physics community
with the counterpart to HERA, which played such a crucial
role together with the Tevatron and LEP machines.  The LHeC 
presents a unique opportunity for utilising
deep inelastic scattering in the
exploration of energy frontier physics, which promises
to rewrite our understanding of high density proton and
nuclear dynamics and structure.

%With the addition of an electron beam, the LHC
%enables the world particle physics community to
%maintain deep inelastic scattering in the
%exploration of energy frontier accelerator
%physics, and it presents a unique perspective to the
%understanding of high density proton and
%nuclear dynamics and structure.

%\subsection{Leptoquarks}
%\input{georges}
%
\newpage
\begin{footnotesize}

%\end{footnotesize}
\newpage
\section*{LHeC Study Group}
\noindent J.L.Abelleira Fernandez$^{16,23}$, 
C.Adolphsen$^{57}$,
P.Adzic$^{74}$, 
A.N.Akay$^{03}$, 
H.Aksakal$^{39}$, 
J.L.Albacete$^{52}$, 
B.Allanach$^{73}$,
S.Alekhin$^{17,54}$, 
P.Allport$^{24}$, 
V.Andreev$^{34}$
% ...

\bigskip{\it\noindent
$^{01}$ Ankara University, Turkey \\
$^{02}$ SANAEM Ankara, Turkey \\
$^{03}$ TOBB University of Economics and Technology, Ankara, Turkey\\
$^{04}$ LAPP, Annecy, France\\
$^{05}$ University of Antwerp, Belgium\\
$^{06}$ University of Birmingham, UK\\
$^{07}$ INFN Bologna, Italy\\
$^{08}$ IIHE, Universit\'e Libre de Bruxelles, Belgium, supported by the FNRS \\
$^{09}$ University of Buenos Aires, Argentina \\
$^{10}$ INFN Gruppo Collegato di Cosenza and Universita della Calabria, Italy \\
$^{11}$ Massachusetts Institute of Technology, Cambridge, USA\\
$^{12}$ Charkow National University, Ukraine\\
$^{13}$ University of Copenhagen, Denmark \\
$^{14}$ Cockcroft Institute, Daresbury, UK\\
$^{15}$ Rutherford Appleton Laboratory, Didcot, UK \\
$^{16}$ CERN, Geneva, Switzerland\\
$^{17}$ DESY, Hamburg and Zeuthen, Germany\\
$^{18}$ University of Hamburg, Germany\\
$^{19}$ Hampton University, USA \\
$^{20}$ University of California, Irvine, USA \\
$^{21}$ University of Jyv\"askyl\"a, Finland\\
$^{22}$ Kastamonu University, Turkey \\
$^{23}$ EPFL, Lausanne, Switzerland\\
$^{24}$ University of Liverpool, UK\\
$^{25}$ University of California, Los Angeles, USA\\
$^{26}$ Lund University, Sweden\\
$^{27}$ University of Wisconsin-Madison, USA\\
$^{28}$ Universidad Aut\'onoma de Madrid, Spain\\
$^{29}$ University of Mainz, Germany\\
$^{30}$ The University of Manchester, UK \\
$^{31}$ INFN Milano, Italy\\
$^{32}$ University of Milano, Italy \\
$^{33}$ University of Montr\'eal, Canada\\
$^{34}$ LPI Moscow, Russia\\
$^{35}$ Tata Institute, Mumbai, India\\
$^{36}$ Jefferson Lab, Newport News, VA 23606, USA \\
$^{37}$ Brookhaven National Laboratory, New York, USA\\
$^{38}$ Columbia University, New York, USA\\
$^{39}$ Nigde University, Turkey\\
$^{40}$ Budker Institute of Nuclear Physics SB RAS, Novosibirsk, 630090 Russia\\
$^{41}$ Orissa University, India\\
$^{42}$ LAL, Orsay, France\\
$^{43}$ Laboratoire de Physique Th\'eorique, Universit\'e Paris XI, Orsay, France \\
$^{44}$ University of Oxford, UK\\
$^{45}$ CPHT, \'Ecole Polytechnique, CNRS, 91128 Palaiseau, France \\
$^{46}$ UPMC University of Paris 06, Facult\'e de Physique, Paris, France \\
$^{47}$ LPNHE University of Paris 06 and 07, CNRS/IN2P3, 75252 Paris, France \\
$^{48}$ Pennsylvania State University, USA\\
$^{49}$ University of Pisa, Italy\\
$^{50}$ University of Pittsburgh, USA\\
$^{51}$ Charles University, Praha, Czech Republic \\
$^{52}$ IPhT Saclay, France\\
$^{53}$ University of Santiago de Compostela, Spain\\
$^{54}$ Serpukhov Institute, Russia\\
$^{55}$ University of Siegen, Germany \\
$^{56}$ University of Southampton, UK \\
$^{57}$ SLAC National Accelerator Laboratory, Stanford, USA\\
$^{58}$ Tokyo Institute of Technology, Japan\\
$^{59}$ University of Torino and INFN Torino, Italy\\
$^{60}$ NTNU, Trondheim, Norway\\
$^{61}$ KEK, Tsukuba, Japan\\
$^{62}$ Uludag University, Turkey\\
$^{63}$ Uppsala University, Sweden  \\
$^{64}$ TRIUMF, Vancouver, Canada \\
$^{65}$ Paul Scherrer Institute, Villigen, Switzerland\\
$^{66}$ National Center for Nuclear Research (NCBJ), Warsaw, Poland \\
$^{67}$ University of Warsaw, Poland\\
$^{68}$ ETH Zurich, Switzerland\\
$^{69}$ University of Zurich, Switzerland \\
$^{70}$ Max Planck Institute Werner Heisenberg, Munich, Germany \\
$^{71}$ QMW University London, United Kingdom \\
$^{72}$ Laboratori Nazionali di Frascati, INFN, Italy \\
$^{73}$ DAMTP, CMS, University of Cambridge, United Kingdom \\
$^{74}$ University of Belgrade, Serbia \\
$^{75}$ RWTH Aachen University, Germany \\
$^{76}$ Instituto Superior T\'{e}cnico, Universidade T\'{e}cnica de Lisboa, Portugal
}

\bigskip{\it\noindent
$^a$ supported by European Research Council grant HotLHC ERC-2011-StG-279579 and\\
MiCinn of Spain grants FPA2008-01177, FPA2009-06867-E and Consolider-Ingenio 2010 CPAN CSD2007-00042,
Xunta de Galicia grant PGIDIT10PXIB206017PR, and FEDER.\\
$^b$ supported by the U.S. Department of Energy,contract DE--AC02--76SF00515.\\
$^c$ supported by the Academy of Finland, project no. 141555.\\
$^d$ supported by the Sloan Foundation,
DOE OJI grant No. DE - SC0002145 and \\
Polish NCN grant DEC-2011/01/B/ST2/03915.\\
$^e$ supported by the Turkish Atomic Energy Authority (TAEK).\\
$^f$ supported by the P2IO consortium.\\
}

\end{footnotesize}
\end{document}